\documentclass[twocolumn,showpacs,amsmath,amssymb,prb]{revtex4}
\usepackage{graphicx}

\usepackage{bm}
\usepackage{color}

\begin{document}

\title{Theory of a Magnetically-Controlled Quantum-Dot Spin Transistor}
\author{Daniel Urban}
\author{Matthias Braun}
\author{J\"urgen K\"onig}
\affiliation{Institut f\"ur Theoretische Physik III, Ruhr-Universit\"at Bochum, 44780 Bochum, Germany}
\date{\today}

\begin{abstract}
We examine transport through a quantum dot coupled to three ferromagnetic leads in the regime of weak tunnel coupling.
A finite source-drain voltage generates a nonequilibrium spin on the otherwise non-magnetic quantum dot.
This spin accumulation leads to magnetoresistance.
A ferromagnetic but current-free base electrode influences the quantum-dot spin via incoherent spin-flip processes and coherent spin precession.
As the dot spin determines the conductance of the device, this allows for a purely magnetic transistor-like operation.
We analyze the effect of both types of processes on the electric current in different geometries.
\end{abstract}


\maketitle

\section{Introduction}
\label{sec:intro}

Electric transport properties depend on magnetic system degrees of freedom as demonstrated by e.g.~the giant- (GMR) and tunnel- (TMR) magneto-resistance effects.
In these it is used that charge currents can be tuned by changing the relative orientation of magnetic moments in magnetic heterostructures.\cite{GMR,TMR} For example, the transmission through an interface between two ferromagnetic electrodes decreases as the relative angle between the electrodes' magnetization directions increases, which is known as the spin valve effect.\cite{reviews}

If a non-magnetic spacer region is inserted between source and drain, the information about the relative spin orientation is mediated between source and drain by spin accumulation in the intermediate region.\cite{jedema:2001,jedema:2003,granularfilms,bernand-mantel:2006,korotkov:1992,zaffalon:2005,kimura:2004}
Scattering at the interfaces may also contribute significantly to transport.\cite{sdips}
Manipulation of the accumulated spin opens the possibility to modify the source-drain current. Such manipulation was suggested to be achieved electrically with the help of spin-orbit coupling as proposed by Datta and Das.\cite{datta-das:1990} It may also be accomplished magnetically, e.g.~by external magnetic fields\cite{pedersen:2005,gorelik:2005,gurvitz:2005,braun:hanle} or by additional leads.\cite{brataas:2001,sanchez:2005,defranceschi:2002}
In addition, in low-dimensional systems the significant charging energy also affects transport, giving rise to the well-known Coulomb staircase\cite{reviews,grabert}, but also affecting the spin-dependence of transport. \cite{rudzinski:2004,braun:2004,braig:2004,wetzels:2005}

In this paper we study a magnetically-controlled quantum-dot spin transistor, 
in which the transport behavior is affected by the interplay of spin 
accumulation and Coulomb charging.
We consider a single-level quantum dot with strong Coulomb interaction that 
is connected to three ferromagnetic leads.
The base lead which is kept charge-current free, so that the source-drain
current is magnetically affected by the base lead only via manipulation of the
accumulated dot spin.

There are two qualitatively different ways in which the current-free base lead affects the quantum-dot spin.
One is that the base electrode offers a channel of spin relaxation.
An electron with, say, spin up tunnels out of the dot and an electron with spin down tunnels in.
Such {\it spin-flip} processes, which are accompanied by a spin but no charge current in the base electrode, reduce the spin accumulation on the dot.
The strength of this relaxation depends on the orientation of the base electrode's magnetization direction relative to the spin accumulation.
This scheme has been proposed to realize a ``spin-flip transistor'' with metallic islands in the absence of Coulomb interaction.\cite{brataas:2001}
In quantum dots, however, there will also be a second contribution, related to {\it spin precession} due to the exchange interaction between the quantum-dot level and ferromagnetic base electrode.\cite{braun:2004}
In general, both types of processes play a role.
The main objective of this paper is to identify and discuss the effect of both of them on the transport characteristics.

\begin{figure}
    \includegraphics[width=.6\columnwidth]{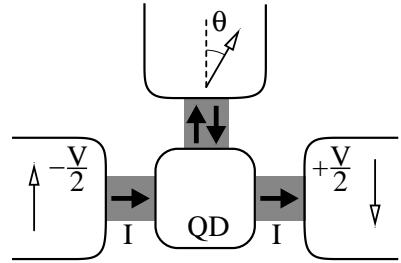}
    \caption{\label{fig:sftrans}
The spin transistor consists of a quantum dot connected to source and drain lead by tunnel contacts. Source and drain have antiparallel magnetizations so that a non-equilibrium spin is accumulated on the dot, which modifies the conductance of the device. A floating base lead can be used to influence the dot state solely by spin currents, i.e., the source-drain conductance can be modified by the alignment of the magnetization alone.}
\end{figure}

In the following Sec.~\ref{sec:model}, we define the model Hamiltonian and
the kinetic equations for the quantum dot's degrees of freedom, and derive the 
conductance of the spin transistor, which is discussed in 
Sec.~\ref{sec:QD-sf-trans}.
In Sec.~\ref{sec:chiral} we discuss the special situation when the three 
magnetizations of the leads are chosen pairwise orthogonal.
In this case, the spin-related resistance change is dominated by the exchange 
effect.
We close with a summary in Sec.~\ref{sec:conclusion}.

\section{Model Hamiltonian, kinetic equations, and current}
\label{sec:model}

We study a single-level quantum dot with contacts to
three ferromagnetic leads [source(left), drain(right),
base(middle)] by tunnel junctions, described by the Hamiltonian
\begin{equation}\label{eq:hamiltonian}
    H   =   H_\text{dot}  +  H_L + H_M + H_R  +  H_T\, .
\end{equation}
The first part $H_\text{dot} = \sum_\sigma \varepsilon\,
c_\sigma^\dagger c_\sigma  +  U n_{\uparrow}n_{\downarrow}$
models the quantum dot as an Anderson impurity with a
spin-degenerate electronic level $\varepsilon$ and charging
energy $U$ for double occupancy.
Each of the three leads is described as a reservoir of
itinerant electrons in thermal equilibrium $ H_r =
\sum_{k\alpha} \varepsilon_{r,k\alpha}^{~}\,
a_{r,k\alpha}^\dagger a^{}_{r,k\alpha} $ with
$r \in \{\text{L},\text{M},\text{R}\}$. Here
$\alpha = +(-)$ denotes the majority(minority)
spin states which have the density of states $\rho_r^\alpha$.
The lead magnetization direction is characterized
by the direction of the magnetization vector
${\mathbf p}_r$, while the strength of the polarization
is given by its magnitude $\left|{\mathbf p}_r\right|
=   (\rho_r^+ - \rho_r^-)/(\rho_r^+ + \rho_r^-)$.

The last part of the Hamiltonian $H_\text{T} =
\sum_{r=L,M,R} H_{T,r}$ connects the four subsystems by spin-conserving tunneling between dot and leads
\begin{equation}\label{eq:HT}
    H_{\text{T},r} =   \sum_{k\alpha,\sigma} V^{r}_{k\alpha,\sigma} a_{r,k\alpha}^\dagger c_\sigma^{~}  \,+  H.c. \,.
\end{equation}
The tunnel matrix elements
$ V_{\alpha,\sigma}^r = t_r \,  \langle\alpha|\,
  \mathrm{e}^{-i \sigma_z \phi_r /2} \, \mathrm{e}^{-i \sigma_y \theta_r /2}
  \,|\sigma\rangle$
consist of the (spin-independent) tunnel amplitude $t_r$, which is a measure
of the barrier height and thickness, times an SU(2) rotation
about the relative polar angles $\theta_r$ and $\phi_r$ between
the lead's magnetization direction and the dot's spin quantization axis.
The tunneling rate for electrons from lead $r$ with spin $\pm$ is quantified by $\Gamma_r^\pm / \hbar = 2 \pi |t_r^2| \rho_r^\pm / \hbar$, and we define
$\Gamma_r = (\Gamma_r^+ + \Gamma_r^-)/2$.
The system considered in this paper extends the
quantum-dot spin valve studied in Ref.~\onlinecite{braun:2004}
by the addition of a third (base) electrode.
We choose this lead to be floating, i.e., to carry no electric charge
current. Instead, it influences the quantum-dot spin, and thus
the source-drain current, only by magnetic interactions.

The state of the quantum dot is characterized by the
probabilities $P_\chi$ to find the dot empty ($\chi=0$),
singly ($\chi=1$), or doubly ($\chi=\text{d}$) occupied,
as well as the average spin $\hbar \mathbf{S}$ with
$\mathbf{S} = (S_x, S_y, S_z)$.
We restrict ourselves to the limit of weak tunnel coupling, for which each 
tunnel event can be uniquely attributed to one lead. 
Therefore the kinetic equations of a quantum dot with three connecting leads 
appears identical to the two-lead case,\cite{braun:2004} whereby the sum over the leads hast to be extended to include also the base (middle) lead.
In lowest order in the tunnel coupling $\Gamma_r$, the kinetic equations read
\begin{widetext}
\begin{equation}
\label{eq:P01d}
  \frac{d}{dt} \! \left(
  \begin{array}{c}
    P_0 \\ P_1 \\ P_{\rm d}
  \end{array}
  \right) \!=\!\!\!\! \sum_{r= L,M,R} \!\! \frac{\Gamma_r}{\hbar} \! \left[ \!\! \left(
  \begin{array}{ccc}
    - 2f^+_r(\varepsilon) & f^-_r(\varepsilon) & 0 \\
    \hphantom{-}2f^+_r(\varepsilon) & - f^-_r(\varepsilon) - f^+_r(\varepsilon+U)
    & \hphantom{-}2f^-_r(\varepsilon+U) \\
    0 & f^+_r(\varepsilon+U) & - 2f^-_r(\varepsilon+U)
  \end{array}
  \right) \! \!\! \left(
  \begin{array}{c}
    P_0 \\ P_1 \\ P_{\rm d}
  \end{array}
  \right)
  \!\! + 2 \! \!\left(
  \begin{array}{c}
    f^-_r(\varepsilon) \\
    - f^-_r(\varepsilon) + f^+_r(\varepsilon+U) \\
    -f^+_r(\varepsilon+U)
  \end{array}
  \right)
    \! {\bf S}\cdot {\bf p}_r  \right] \! ,
\end{equation}
\end{widetext}
for the occupation probabilities of the different charge states, while the spin degrees of freedom are determined by
\begin{equation}
\label{eq:S}
  \frac{d\mathbf S}{dt} = \sum_{r=L,M,R}
  \frac{\Gamma_r}{\hbar}
     \left( f^+_{r}(\varepsilon) P_0 + \frac{-f^-_{r}(\varepsilon)+f^+_{r}(\varepsilon+U)}{2} P_1 - f^-_{r}(\varepsilon+U)P_{\rm d}  \right) {\mathbf p}_r
 - \frac{\Gamma_r}{\hbar} \left[ f_r^-(\varepsilon)+f_r^+(\varepsilon +U) \right] \mathbf{S}
 + {\mathbf S} \times {\mathbf B}_r     \,,
\end{equation}
where $f_r^+(\omega)$ labels the Fermi function of lead $r$, and $f_r^-(\omega) = 1-f_r^+(\omega)$.
The kinetic equation (\ref{eq:S}) contains a term describing coherent spin 
precession about an effective magnetic field, mediated by tunneling between 
dot and leads
\begin{equation}\label{eq:exchfield}
  {\mathbf B}_r = \frac{1}{\hbar} {\mathbf p}_r \frac{1}{\pi} \int'  d\omega \, \Gamma_r(\omega) \left( \frac{f^+_r(\omega)}{\hbar\omega-\varepsilon-U} + \frac{f^-_r(\omega)}{\hbar\omega-\varepsilon} \right)   ,\,
\end{equation}
where the prime on the integral symbolizes Cauchy's principal value. In this paper we consider the case of flat bands such that $\Gamma_r$ is independent of $\omega$ and the exchange field Eq.~(\ref{eq:exchfield}) vanishes in the absence of charging energy. 

By solving the system of kinetic equations (\ref{eq:P01d}) and (\ref{eq:S}) in the stationary limit (under the constraint of probability normalization)
we obtain the charge and spin occupation probabilities. The stationary charge current into lead $r$ is then given by
\begin{eqnarray}
\label{eq:current}
I_r =&&\!\!\!\!\! \frac{2(-e)\Gamma_{r}}{\hbar} \Bigl(
     f^+_{r}(\varepsilon) P_0
     + \frac{-f^-_{r}(\varepsilon) +f^+_{r}(\varepsilon+U)}{2} P_1\nonumber\\
     &&\!\!\!\!\!- f^-_{r}(\varepsilon+U) P_{\rm d}
     -\left[ f^-_r(\varepsilon) + f^+_r(\varepsilon+U) \right]
     {\mathbf S}\cdot {\mathbf p}_r
     \Bigr)   \,.
\end{eqnarray}

We see that the quantum-dot charge and spin degrees
of freedom are coupled to each other, and both enter
the expression for the stationary current.
In particular, an accumulation of spin in the quantum
dot due to a finite source-drain voltage will reduce
electric transport, which constitutes the spin valve
effect.

The goal of this paper is to study how transport through this spin valve can be controlled in a purely magnetic way by means of the third lead.
In order for this control to be purely magnetic we keep
the base electrode floating, i.e., it does not carry
any net charge current, $I_M=0$.

The base lead influences the quantum-dot spin in two qualitatively different 
ways.
First, the base electrode can act as (incoherent) sink for spin currents, thus offering a channel of spin relaxation.
Second, the exchange field originating from the tunnel contact can give rise to coherent precession of the quantum-dot spin.
Both effects could lead to a transistorlike behavior as discussed
in the next section.

\section{Quantum-Dot Spin Transistor}
\label{sec:QD-sf-trans}

In this section we concentrate on the quantum-dot
spin transistor as shown in Fig.~\ref{fig:sftrans}.
Source and drain electrodes are magnetized antiparallel
to each other. This maximizes the magnitude of the
accumulated spin on the quantum dot. The magnetization
direction of the base electrode encloses an angle $\theta$
with the source electrode.
We are interested in the dependence of the source-drain
current on the angle $\theta$.

For simplicity we assume equal tunnel couplings $\Gamma_\text{L}=\Gamma_\text{R}=\Gamma$
and polarizations $p_L=p_R=p$ of the source and drain electrodes.
Furthermore, we apply the transport voltage $V$
symmetrically, $V_\text{L}=-V_\text{R}=V/2$, and
focus on the linear-response conductance
$G=\partial I/\partial V\big|_{V=0}$ for
source-drain voltages $e V \ll k_B T$ much
smaller than the temperature.

In this case, the conductance can be expressed in
terms of typical time scales that determine the
quantum-dot charge and spin dynamics.
The characteristic time scale of charge transport,
$\tau_c$, is the lifetime of the
singly occupied charge state, limited by tunneling
out of the dot or tunneling in of a second electron
to or from the source or drain lead,
\begin{equation}
  \frac{1}{\tau_c}  = \frac{2\Gamma}{\hbar}
  \left[f^-(\varepsilon)+f^+(\varepsilon +U)\right] \, .
\end{equation}
This time scale directly reflects the electrical current through the 
quantum dot: in the absence of a source and drain lead polarization we 
would obtain the linear conductance
$G=\left.\partial I/\partial V \right|_{V=0}$ as
\begin{equation}\label{eq:G0}
    G_0 =   \frac{e^2}{\hbar} \frac{1}{k_B T} \frac{P_1^{(0)}}{\tau_c}\,.
\end{equation}
The conductance is directly proportional to $1/\tau_c$ times
the equilibrium probability to find the dot singly occupied,
$P_1^{(0)} = 2 f^+(\varepsilon) f^-(\varepsilon +U)
/ [f^+(\varepsilon) + f^-(\varepsilon +U)]$.

The time scale for spin transfer or spin coherence
is somewhat more subtle. Tunneling processes from or to
ferromagnetic leads generate a finite spin accumulation
as well as offering relaxation channels, which limit the spin
lifetime.
In Ref.~\onlinecite{braun:2004} we chose to call the first
and second terms in Eq.~(\ref{eq:S}) the accumulation and relaxation
terms, respectively.
For the source and drain leads we keep this interpretation.
The separation of the tunneling processes into spin accumulation
and relaxation terms is to some degree arbitrary.
To define a proper spin lifetime boundary conditions need to be specified.
In the case of the base lead, the condition $I_M=0$ allows us
to rewrite the contribution of the middle (base) lead $M$ to the spin
kinetic equation (\ref{eq:S}) as
{
\begin{eqnarray}\label{eq:SM}
  \frac{d\mathbf S}{dt}  \bigg|_M \!\!\!\!\!= -\frac{\Gamma_M}{\hbar}
  \left[ f^-(\varepsilon)+f^+(\varepsilon +U) \right]
  \left[ {\mathbf S} - ( {\mathbf S} \cdot {\mathbf p}_M ){\mathbf p}_M \right] \nonumber\\
  +{\mathbf S} \times {\mathbf B}_M \, .
\end{eqnarray}
}
By removing the spin-accumulation term from the kinetic equation we observe that the damping becomes anisotropic.
Putting these pieces together leads to the definition of a spin lifetime
\begin{equation}
  \label{eq:tauspara}
  \frac{1}{\tau_{s_\parallel}}
  = \frac{2\Gamma +(1-p_\text{M}^2)\Gamma_\text{M}}{\hbar}
  \left[f^-(\varepsilon)+f^+(\varepsilon +U)\right]
  \, ,
\end{equation}
for the case when the magnetization of the base lead is
parallel to source and drain leads. For orthogonal alignment
the spin lifetime becomes
\begin{equation}
  \frac{1}{\tau_{s_\perp}}    =
  \frac{2\Gamma +\Gamma_M}{\hbar}
  \left[f^-(\varepsilon)+f^+(\varepsilon +U)\right] \,.
\end{equation}

If the source and drain have a finite polarization $p$,
an average spin accumulates on the dot, giving rise
to a magnetoresistive effect which reduces the
conductance proportional to $p^2$.
In the case of parallel source and base magnetizations
($\theta=0$) the base magnetization is also parallel to
the accumulated spin. Due to this collinearity the
precession term in Eq.~(\ref{eq:S}) vanishes and the
conductance (\ref{eq:G0}) is reduced to
\begin{eqnarray}
    \label{eq:Gpara}
    G_\parallel &=& G_0 \left(1-\frac{\tau_{s_\parallel}}{\tau_c} p^2 \right)\,.
\end{eqnarray}
The characteristic charge transport time $\tau_c$ is
independent of the base lead as it is floating.
In contrast the base lead may carry spin currents and thus reduce the spin lifetime $\tau_s<\tau_c$.

In the case of perpendicular magnetization
alignment $\theta=\pi/2$, the
intrinsic coherent spin precession also becomes important.
In the stationary situation the electrical currents
through source and drain interface are equal,
therefore we may focus on the dot-drain interface only.
This interface can be seen as a tunnel magnetoresistance
element of its own, i.e., its conductance depends on the relative
angle between dot spin ${\mathbf S}$ and lead magnetization
${\mathbf p_R}$. Since the exchange field originating from
the base lead modifies the dot-spin direction, it also
modifies the conductance of the device.\cite{note1}
In the orthogonal magnetization alignment the spin
precession effect is maximally pronounced, and the
magnetoresistance of the total device is reduced by the
factor $1/(B_\text{M}^2\tau_{s_\perp}^2 + 1)$ to
\begin{eqnarray}
    \label{eq:Gortho}
    G_\perp &=& G_0 \left(1-\frac{\tau_{s_\perp}}{\tau_c} p^2 \frac{1}{B_\text{M}^2 \tau_{s_\perp}^2 + 1}\right)     .
\end{eqnarray}

\begin{figure}
    \includegraphics[width=\columnwidth]{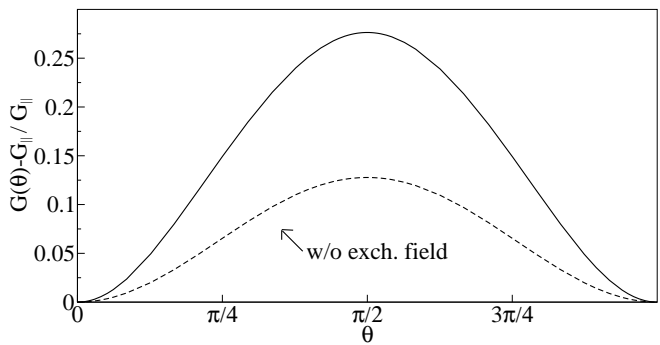}
    \caption{\label{fig:angdepTMR}
    Angular dependence of the normalized conductance change $\Delta = (G(\theta)-G_\parallel)/G_\parallel$ of the spin transistor (solid line). Comparison to the case where the exchange field was artificially set to zero (dashed) reveals that the transistor effect is significantly enhanced by exchange interaction. 
The parameters are $\varepsilon = 0.6\,U$, $p_L=p_\text{M}=p_R=0.6$, $\Gamma_\text{L}=\Gamma_\text{M}=\Gamma_\text{R}$, and $U=10\,k_BT$.}
\end{figure}

In Fig.~\ref{fig:angdepTMR} we plot the normalized conductance change
$\Delta = [G(\theta)-G_\parallel]/G_\parallel$
of the transistor structure as a function of the base
lead magnetization direction. For $\theta=\pi/2$, the
device conductance is maximal, since in this alignment
both the spin relaxation due to the base lead and the
spin-precession effect maximally suppress the
magnetoresistance caused by spin accumulation.
By comparison with the dashed curve, for which the exchange
field was set to zero manually, it can be seen that the
transistor shows significant influence of the exchange
interaction which is caused by electron-electron interaction.

\begin{figure}
    \includegraphics[width=1\columnwidth]{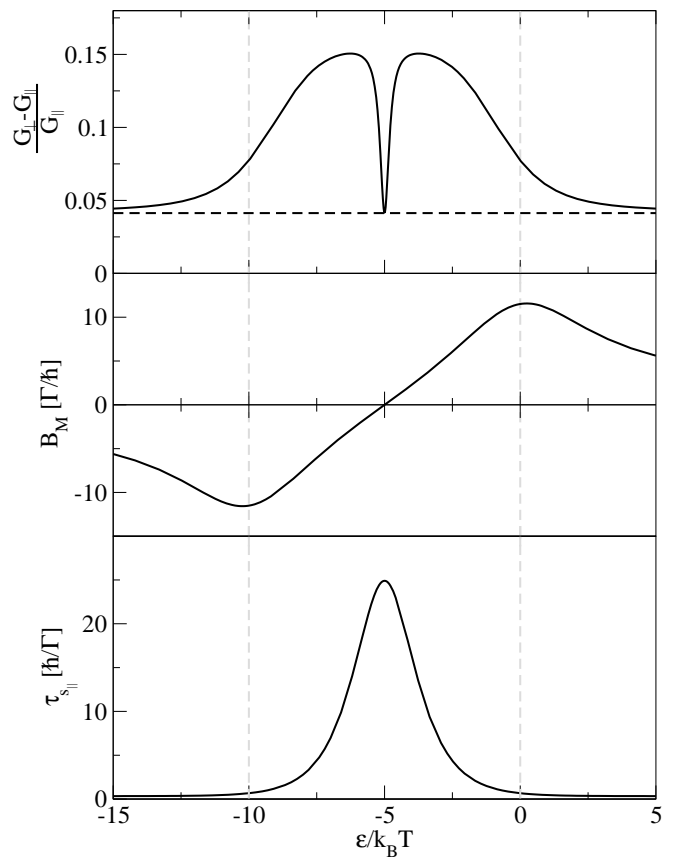}
    \caption{\label{fig:VdepTMR}
    (a) Gate voltage dependence of $\Delta = (G(\theta)-G_\parallel)/G_\parallel$ with (solid) and without (dashed) exchange field for $\theta=\pi/2$. The level position is at $\varepsilon=0$, and charging energy is $U=10\,k_BT$.
    (b) Absolute value of the exchange field of the middle lead for $p_M=1$.
    (c) Spin lifetime in the parallel case $\tau_{s_\parallel}$. The product $B_M\tau_{s_\parallel}$ determines the conductance change.
    Other parameters as in Fig.~\ref{fig:angdepTMR}.
    }
\end{figure}

In Fig.~\ref{fig:VdepTMR}(a), the maximum value of $\Delta(\theta=\pi/2)$ is plotted as a function of gate voltage.
$\Delta$ shows strong variations, which arise only due to the
gate-voltage dependence of the exchange field from the base lead [Fig.~\ref{fig:VdepTMR}(b)].
Without the exchange interaction, no gate-voltage dependence
is expected (see dashed line). Due to its strong gate-voltage dependence,
the exchange interaction contribution can be separated
from the influence of the anisotropic spin-flip relaxation,
which does not depend as strongly on the gate voltage.

\section{Orthogonal Magnetization}
\label{sec:chiral}

In the quantum-dot spin transistor geometry shown
in Fig.~\ref{fig:sftrans} and discussed in the previous
section, both spin-flip and spin-precession processes
contribute at the same time. In the last part of this
paper, we want to isolate the exchange-field contribution
in a purely magnetic way.

\begin{figure}
    \includegraphics[width=.6\columnwidth]{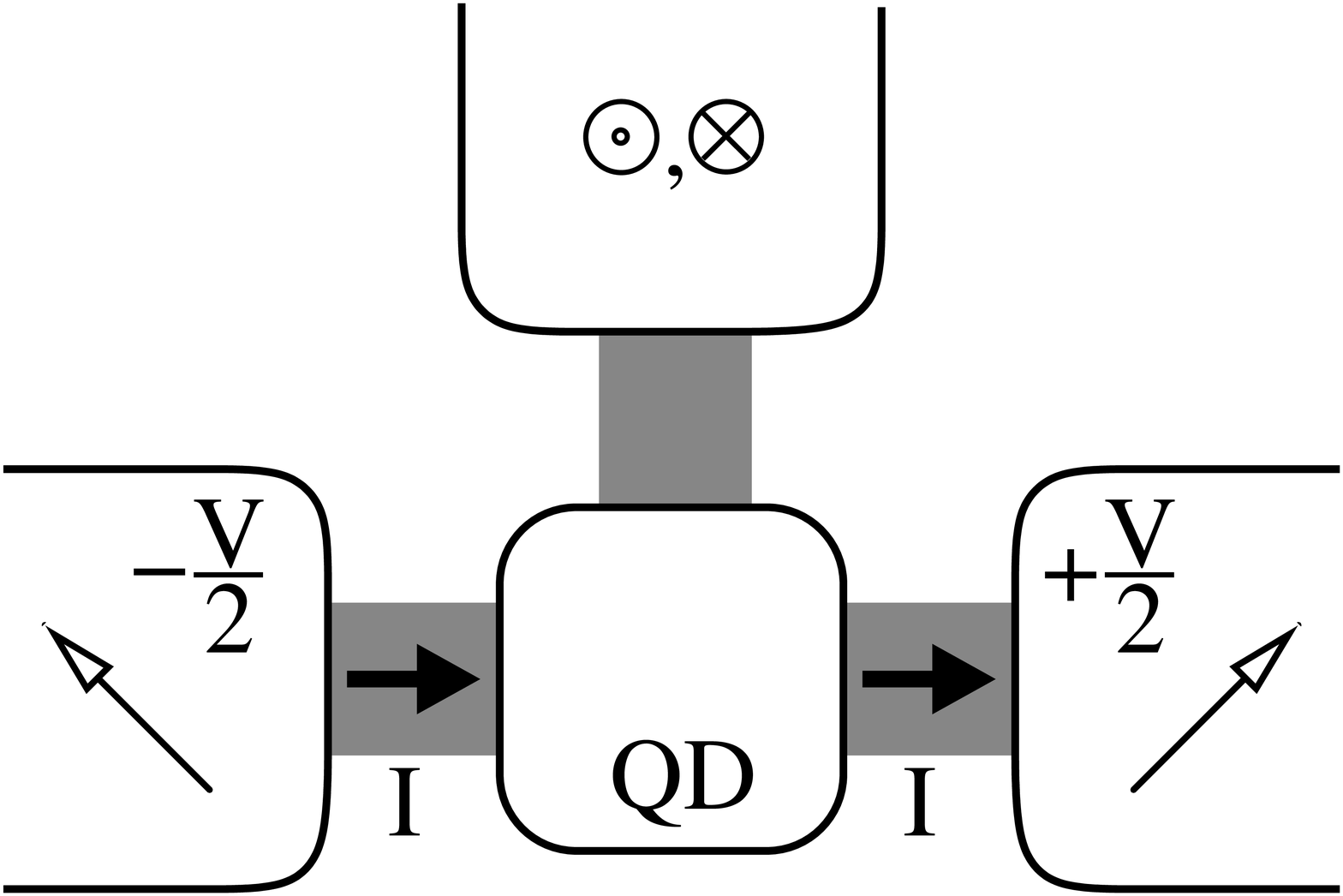}
    \caption{\label{fig:chiral}
    The two realizations of a system with pairwise orthogonally magnetized leads are connected by reversal of the base magnetization. Their different symmetry can be reflected in the symmetry of the conductances.}
\includegraphics[width=1\columnwidth]{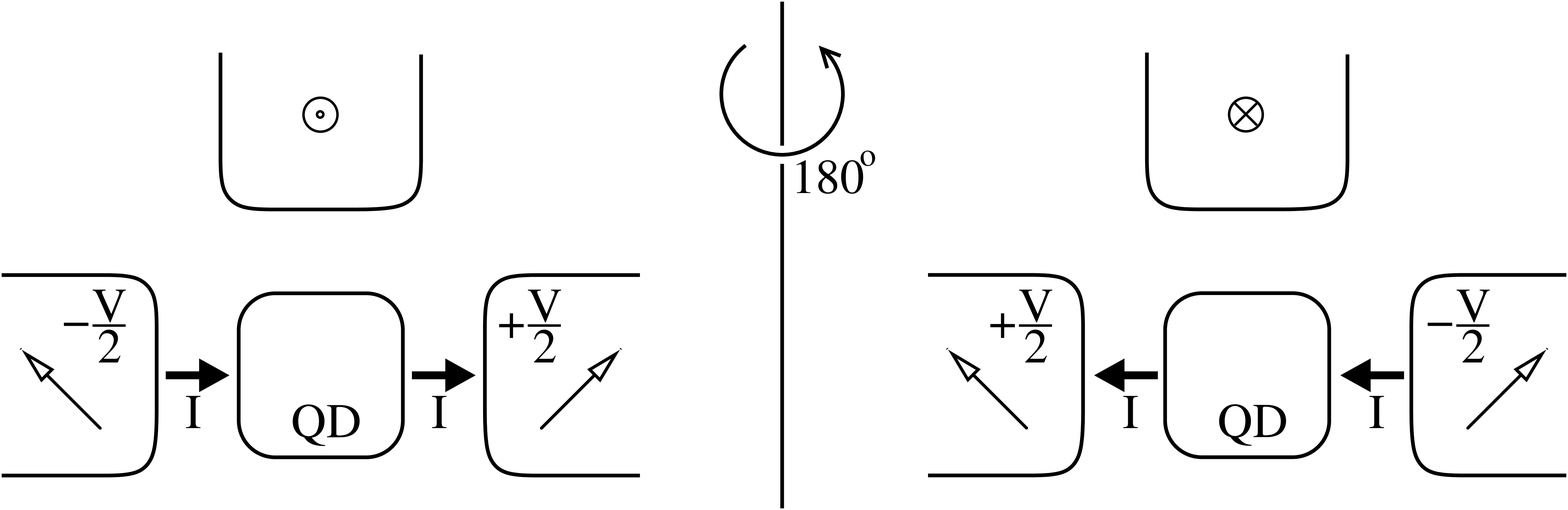}
\caption{
\label{fig:symmarg}
A rotation
shows that the right-handed system is symmetric to the left-handed system with reversed current and voltage.}
\end{figure}

For this, we consider a geometry of the quantum-dot spin
transistor with pairwise orthogonal magnetization
directions in the non-linear response regime (see
Fig.~\ref{fig:chiral}).
There are two realizations of such a system, defined by $ {\mathbf p}_L \cdot ({\mathbf p}_M \times {\mathbf p}_R) \lessgtr 0$. They are related by reversal of the base magnetization or reversal of the voltage (and thus current directions). For these systems the anisotropic damping is equal, as it gives rise only to factors $1-{\mathbf p}_M^2$. Any difference in the transport behavior of the two systems can thus be attributed entirely to exchange effects.

The exchange field now always has contributions from
all three leads, as source and drain contributions never
cancel each other. Correspondingly, the axis and angle of precession
of the dot spin depends on all three lead potentials and
is not easily visualized, in particular in the case of
finite bias voltages.

In the linear-response regime the conductances of the left- (LHS) and 
right-handed system (RHS) are equal to each other.
This is a consequence of the symmetry of the system for equal source and drain parameters ($p_L=p_R$, $\Gamma_L=\Gamma_R$) (see Fig.~\ref{fig:symmarg}).

\begin{figure}
\begin{center}
\includegraphics[width=1\columnwidth]{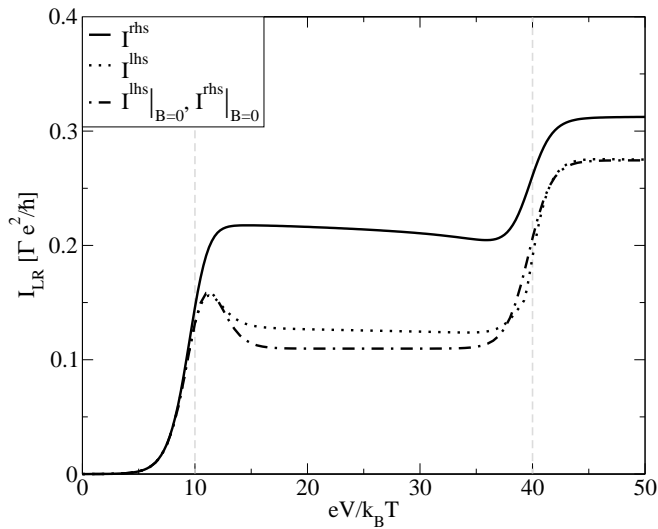}
\caption{
\label{fig:ch-fr-current}
Current-voltage characteristic for the orthogonal configuration with $p_L=p_R=p_\text{M}=0.6$, $\Gamma_\text{L}=\Gamma_\text{R}=\Gamma_\text{M}=\Gamma$, $\varepsilon=10\,k_BT$, $U=30\,k_BT$.}
\end{center}
\end{figure}

In order to observe a difference in the conductances of the LHS and RHS, the symmetry of source and drain has to be broken, which is best done by application of finite bias voltages (Fig.~\ref{fig:ch-fr-current}). If the exchange fields are set to zero manually in the kinetic equations, a mechanism similar to the one described in Ref.~\onlinecite{braun:2004} leads, for $p\rightarrow1$, to complete spin blockade between $\varepsilon$ and $\varepsilon +U$: driven by the current, the accumulated spin tends to align antiparallel to the drain electrode and thus blocks transport (dot-dashed line).
In the presence of an exchange field this blockade is lifted due to precession of the accumulated spin. This process differs for the left- and right-handed systems due to the different precession directions, which are reflected in different conductances.

\section{Conclusion}
\label{sec:conclusion}

We analyzed electron transport through a magnetically controlled quantum-dot spin transistor in the regime of weak dot-lead tunnel coupling. The presence of Coulomb interaction on the dot gives rise to an exchange interaction of the accumulated dot spin with the ferromagnetic leads, giving rise to spin precession.
Furthermore, the tunnel coupling to a current-free ferromagnetic base electrode
leads to an anisotropic spin relaxation: dot spins oriented along this lead's 
magnetization experience a weakened damping, i.e., an increased spin lifetime 
relative to other spin orientations.

These two effects allow for a purely magnetic control of the source-drain 
current.
We demonstrated this for two setups.
In the first, with source and drain magnetized antiparallel to each other, we
found a dependence of the source-drain voltage on the magnetization direction 
of the base lead.
The influence of the exchange field becomes clearly visible in the gate-voltage
dependence of the conductances, allowing for the possibility of separating the two 
effects in experiments.
As a second possibility to isolate the exchange-field contribution we 
propose a setup with all three leads' magnetizations being pairwise orthogonal 
to each other.
In this case the difference of the currents for the left- and right-handed 
systems is purely due to the exchange field.
  
\begin{acknowledgments}
This work was supported by DFG under SFB 491 and GRK 726. J.K.~acknowledges
the hospitality of the CAS Oslo.
\end{acknowledgments}

\appendix
\section{Exchange field and spin-mixing conductance}

The accumulated spin $\bf S$ on the quantum dot is coupled to the magnetization of the ferromagnetic lead via virtual tunnel processes. This coupling leads to a precession of the dot spin around the lead magnetization direction, described by the exchange field $\bf B $ in Eq.~(\ref{eq:exchfield}).
In terms of angular momentum transfer the precession is equivalent to a spin current transverse to both lead magnetization and dot spin-accumulation direction. The transversality is directly reflected in the cross-product structure $(\bf S \times \bf B)$ in Eq.~(\ref{eq:S}).

For noninteracting systems, spin transport though interfaces has been studied extensively in terms of a scattering-wave approach.\cite{brataas:2006} Brataas {\it et al.}~found\cite{brataas:2000,brataas:2001} that spin transport can be characterized by three parameters: the conductance for spin-up electrons, that for spin-down electrons, and the complex spin-mixing conductance
\begin{eqnarray}\label{app:spinmixing}
G^{\uparrow\downarrow}(\omega)=\frac{e}{h}\left[1- r_{\uparrow\uparrow}(r_{\downarrow\downarrow})^\star\right],
\end{eqnarray}
where $r_{\sigma\sigma}$ are the reflection amplitudes of electrons with spin $\sigma$ and energy $\omega$.
Electrons entering the junction from the normal side and being reflected at 
the interface acquire a spin-dependent phase shift, equivalent to a rotation 
of the spin state about the magnetization direction of the ferromagnet.
This mechanism, also discussed in the context of singlet-triplet mixing in ferromagnet-superconductor heterostructures,\cite{FMSCheterostructures}
is described by the imaginary part of the spin-mixing conductance and leads to 
a transverse component of the spin current through an interface.

We now show how the exchange field that we calculate in terms of Green's functions can, for the noninteracting limit $U=0$, be matched to the imaginary part of the spin-mixing conductance. For this we consider an electron arriving from the ferromagnet, which scatters back at the ferromagnet--quantum-dot interface.\cite{note2} By means of the Fisher-Lee relations,\cite{fisher:1981,meir:1992} we can relate the reflection amplitude
\begin{eqnarray}
  r_{\sigma\sigma}(\omega) = 
  -1+i \Gamma_\sigma(\omega) G_{\sigma\sigma}^{\rm ret}(\omega)
\end{eqnarray}
to the retarded Green's function $G^{\rm ret}(\omega)$ of the quantum dot and the spin-dependent tunnel coupling $\Gamma_\sigma$, which yields
\begin{eqnarray}
    G^{\uparrow\downarrow}(\omega)
            &=& \frac{i e}{h} \left(\,\Gamma_\uparrow G_{\uparrow\uparrow}^{\rm ret} -\Gamma_\downarrow G_{\downarrow\downarrow}^{\rm adv}\, \right)
       \nonumber\\
            &=& \frac{e}{h} (\Gamma_\uparrow+\Gamma_\downarrow) \left(
      - {\rm Im} \, G^{\rm ret} + i p \, {\rm Re} \, G^{\rm ret} \right)
\end{eqnarray}
plus terms of higher order in $\Gamma$.
In the last line, we dropped the spin indices on the Green's functions, as in the limit of zeroth order in $\Gamma$, the diagonal Green's functions $G^{\rm ret}_{\sigma\sigma}(\omega)=1/(\omega-\varepsilon+i0^+)$ are independent of spin.\cite{note3}

At this point, we can identify two conceptually different contributions to the spin mixing conductance: the real part of the conductance is proportional to the spectral density of the quantum dot, associated with a real particle 
transfer between lead and dot. On the other hand, the imaginary part of the mixing conductance is proportional to the polarization $p=(\Gamma_\uparrow-\Gamma_\downarrow)/(\Gamma_\uparrow+\Gamma_\downarrow)$ and to ${\rm Re} \, G^\text{ret}$. To get the total transfer of angular momentum between dot and lead, we need to integrate over all frequencies $\omega$, which gives rise to a principal value integral with integrand $1/(\omega-\epsilon)$, in agreement with a full Green's function formulation of the spin current through a tunnel 
barrier.\cite{braun:2005} 

Eventually, we find the resulting integral equal to the expression of the 
exchange field  $\mathbf B_r$ in Eq.~(\ref{eq:exchfield}) in the noninteracting limit $U=0$,
\begin{equation}
    \int d\omega  \,\text{Im}\,G^{\uparrow\downarrow}(\omega)  =  2e \, |\mathbf B_r|
\end{equation}

This result complements the scattering picture,\cite{brataas:2000,brataas:2001,brataas:2006} within which the spin-mixing conductance emerges from the shape of the interface barriers. Such structural aspects are not accessible within the Green's function formalism employed in this paper, since the underlying Hamiltonian describes high and narrow barriers. The scattering wave approach, on the other hand, is unable to describe many-particle effects such as the exchange field. While both contributions are present in the spin-mixing conductance, it depends on the specific system which one dominates.


\begin{thebibliography}{99}
\bibitem{note1}
For $e V \ll k_B T$, the exchange interactions originating
from source and drain cancel due to the antiparallel
alignment (${\mathbf B}_L+{\mathbf B}_R=0$),
so that only the exchange interaction from the base lead ($M$)
remains.
\bibitem{note2}
This approach requires the vanishing of the dot Green's functions nondiagonal in spin space, which is always satisfied in the weak-coupling limit.
\bibitem{note3}
It is worth mentioning that the retarded and advanced zeroth-order Green's functions of the dot electrons are formally equal to the free Green's functions, even though this is not the same limit. The zeroth-order approximation takes into account fully the nonequilibrium situation of the system. This  point becomes more transparent by considering that the off-diagonal greater and lesser Green's functions vanish in the free limit, while in zeroth-order tunnel coupling $\Gamma$ they do not, but rather describe nonequilibrium spin accumulation on the quantum dot.
\end{thebibliography}
\end{document}